\documentclass[twocolumn,showpacs,superscriptaddress]{revtex4}
\usepackage{amssymb}
\usepackage{mathrsfs}
\usepackage{graphicx}
\usepackage{float}
\usepackage{latexsym}
\usepackage{bm}

\begin{document}

\title{Ferromagnetic to antiferromagnetic transition of one-dimensional
spinor Bose gases with spin-orbit coupling}
\author{Xing Chen}
\email{chenxing@iphy.ac.cn}
\affiliation{Beijing National Laboratory for Condensed Matter Physics, Institute of
Physics, Chinese Academy of Sciences, Beijing 100190, China}
\author{Haiping Hu}
\affiliation{Beijing National Laboratory for Condensed Matter Physics, Institute of
Physics, Chinese Academy of Sciences, Beijing 100190, China}
\author{Yuzhu Jiang}
\affiliation{Beijing Computational Science Research Center, Beijing 100084, China}
\author{Shu Chen}
\affiliation{Beijing National Laboratory for Condensed Matter Physics, Institute of
Physics, Chinese Academy of Sciences, Beijing 100190, China}
\date{ \today}

\begin{abstract}
We investigate the interacting two-component bosonic gases with spin-orbit (SO) coupling 
in one dimension. Through a gauge transformation, the effect of SO coupling is incorporated
into a spin-dependent twisted boundary condition. We solve 
the SO coupled system analytically by using the BA method. Our result shows that the
SO coupling can influence the eigenenergy in a periodical pattern. The
interplay between interaction and SO coupling may induce the energy level
crossings for the lowest energy spectrum, which leads to a transition from the
ferromagnetic to antiferromagnetic state.
\end{abstract}

\maketitle
\date{}

\section{Introduction}

The experimental success in manipulating cold atoms in effective
one-dimensional (1D) waveguides has deepened our understanding of the
properties of the low-dimensional quantum gases \cite%
{Paredes,Toshiya,Cazalilla,GuanXW}. Meanwhile, studies of synthetic gauge
field in cold atom systems have also made great progresses: pioneering
experiments of NIST group have generated the effective magnetic fields,
electric fields and spin-orbit (SO) coupling in ultracold Bose gases \cite%
{Y.-J. Lin}. The SO-coupled Fermionic gases have also been realized recently
\cite{Zhang,Zwierlein}. Intriguing phenomena in condensed matter physics, 
such as quantum spin Hall effect and topological insulators \cite{X. L. Qi,M. Z. Hasan}, 
where electrons play the elemental role in these physical systems, 
are revealed in the SO coupling systems. The realization of SO coupling in
cold bosonic systems opens a completely new avenue for studying the physics of
Abelian or non-Abelian gauge potentials beyond the traditional condensed matter physics.

Many theoretical researches have revealed  interesting
phenomena for SO coupled bosonic systems. For example, a single plane wave
phase or a stripe phase in spin-1/2 Bose-Einstein condensates (BECs) with SO
coupling has been predicted depending on the intraspecies interaction larger
or smaller than that of interspecies \cite{C. Wang,Ho}. The collective modes
\cite{W. Zheng}, stability  of BECs with SO coupling \cite{Ozawa,BiaoWu}, and
the phases in the presence of harmonic traps and rotation \cite{Wu, X.-Q. Xu, J. Radi'c, B.
Ramachandhran, S. Sinha,ZFXu} have also been studied. Most of these investigations restricted on
mean-field approximation in the weakly interacting regime. However, mean-field
theory fails in the strong interaction limit. In order to get a complete physical picture 
of SO coupled system, the exact solution for SO coupled cold atom system is highly desirable.

In this paper, we solve analytically the SO coupled spin-1/2 bosonic system
in a ring trap by the Bethe-ansatz (BA) method. In this system, SO coupling affect the 
eigenenergy periodically. A pioneering research proposed
to realize the spatially periodic Raman coupling for a two component Bose-Einstein condensates
by using two intersecting laser beam, which provides a platform to investigate these system
experimentally \cite{J. Higbie}. The SO coupling
brings new physics to Bosonic system, particularly in the strongly
interacting regime. In the absence of SO coupling, the ground state of the
 spin-1/2 bosonic system is the ferromagnetic state by solving BA equations
\cite{Y. Q. Li,Guan Xiwen,Yajiang Hao,Fuchs}. By adding the SO coupling,
we find that the competition between the SO coupling and interaction produce
energy level crossings for the lowest energy spectrum in the strongly interacting regime.
The ground state of the system may change from a ferromagnetic state to
an antiferromagnetic state.

The present paper is organized as follows. In Section II, we introduce
the model and solve it by using BA method. By employing  a rather general transformation, 
SO coupling effect could be transformed to the twisted boundary condition. 
The eigenenergy is got by solving the BA equtions and the energy
level crossings are shown. In Section III, we study the BA equations
 in the strongly interacting regime and demonstrate that the
antiferromagnetic state can be the ground state. A summary is given in Section IV.

\section{Model and solution}

We consider a two-component bosonic gas confined in a 1D ring trap in the
presence of SO coupling with the Hamiltonian given by $\hat{H}=\hat{H}_{0}+%
\hat{H}_{\text{int}}$ with
\begin{equation}
\hat{H}_{0}=\int dx\hat{\Psi}^{\dag }\left( x\right) \left( -\frac{\hbar ^{2}%
}{2m}\frac{\partial ^{2}}{\partial x^{2}}-2i\alpha \sigma _{z}\frac{\partial
}{\partial x}\right) \hat{\Psi}\left( x\right) ,  \label{Ham1}
\end{equation}%
where $\hat{\Psi}^{\dag }=(\hat{\Psi}_{\uparrow }^{\dag },\hat{\Psi}%
_{\downarrow }^{\dag })$ represents two internal states of bosonic atoms, $%
\alpha $ denotes the spin-orbit coupling strength and $\sigma _{z}$ is the
Pauli matrix. Here, we consider a quasi-one-dimensional situation with the
transverse motion tightly confined in its ground state. The ring trap
enforces the periodic boundary condition $\hat{\Psi}_{\sigma }\left(
x\right) =\hat{\Psi}_{\sigma }\left( x+L\right) $. The interaction term is
generally represented as $\hat{H}_{\text{int}}=\int dx(g_{1}\hat{n}%
_{\uparrow }^{2}+g_{2}\hat{n}_{\downarrow }^{2}+2g_{12}\hat{n}_{\uparrow }%
\hat{n}_{\downarrow })$, where $g_{1,2}$ and $g_{12}$ denote the strengths
of intraspecies and interspecies interaction, which are experimentally
tunable. In this work, we shall focus on the case with spin-independent
interaction, i.e., $g_{1}=g_{2}=g_{12}=c$, for which interaction term can be
represented as $\hat{H}_{\text{int}}=c\int dx\hat{n}^{2}$ with $\hat{n}=\hat{%
n}_{\uparrow }+\hat{n}_{\downarrow }$. We shall set $\frac{\hbar ^{2}}{2m}=1$
in the following text for convenience.

It is difficult to solve the Hamiltonian with spin and momentum coupled
together. Using a rather general transformation \cite{Shun Uchino}
\begin{equation}
\hat{\Psi}_{\sigma }\left( x\right) =\hat{\Phi}_{\sigma }\left( x\right)
e^{-i\alpha x \sigma _{z}},  \label{TRS}
\end{equation}%
the Hamiltonian is rewritten as follows,
\begin{equation}
\hat{H}_{0}=-\sum_{\sigma }\int dx \hat{\Phi}_{\sigma }^{\dag }\left(
x\right) \frac{\partial ^{2}}{\partial x^{2}}\hat{\Phi}_{\sigma }\left(
x\right) -N\alpha ^{2},  \label{HamB}
\end{equation}%
and the form of $\hat{H}_{\text{int}}=c\int dx\hat{n}^{2}$ is invariant with
$\hat{n}=\hat{\Phi}^{\dag }\hat{\Phi}$ and $\hat{\Phi}^{\dag }=(\hat{\Phi}%
_{\uparrow }^{\dag },\hat{\Phi}_{\downarrow }^{\dag })$. The operator $\hat{%
\Phi}_{\sigma }\left( x\right) $ and $\hat{\Phi}_{\sigma }^{\dag }\left(
x^{\prime }\right) $ also satisfy commutation relation $[\hat{\Phi}_{\sigma
}^{\dag }\left( x\right) ,\hat{\Phi}_{\sigma ^{\prime }}\left( x\right)
]=\delta _{\sigma \sigma ^{\prime }}\delta (x-x^{\prime })$. Meanwhile, the
total momentum now is represented as $\hat{K}=\sum_{\sigma }\int dx\hat{\Phi}%
_{\sigma }^{\dag }\left( x\right) \left( -i\frac{\partial }{\partial x}%
-\alpha \sigma _{z}\right) \hat{\Phi}_{\sigma }\left( x\right) $ which is
spin-dependent. The Schr\"{o}dinger equation is $\hat{H}|\Phi \rangle
=E|\Phi \rangle $, where the wave function is given by
\begin{equation}
|\Phi \rangle =\sum_{\kappa }\int \phi _{\kappa }\left( x_{1}...x_{N}\right)
\prod_{i=1...N}\hat{\Phi}_{\sigma _{i}}^{+}\left( x_{i}\right)
dx_{i}\left\vert 0\right\rangle ,  \label{WFS}
\end{equation}%
here $\kappa $ denotes $\sigma _{1},\sigma _{2}...,\sigma _{N}$
corresponding to the spin index for different particles. By applying the
wave function to the Schr\"{o}dinger equation, we get
\begin{equation}
H\phi _{\kappa }\left( x_{1},...,x_{N}\right) =E\phi _{\kappa }\left(
x_{1},...,x_{N}\right)  \label{SchE}
\end{equation}%
with
\begin{equation}
H=-\sum_{i=1}^{N}\frac{\partial ^{2}}{\partial x_{i}^{2}}+2c\sum_{i<j}\delta
\left( x_{i}-x_{j}\right) -\alpha ^{2}N,  \label{HamF}
\end{equation}%
It is worth mentioning that SO effect is not omitting but transformed to
the spin-dependent twisted boundary condition. Explicitly, the periodic
boundary condition $\hat{\Psi}_{\sigma }\left( x+L\right) = \hat{\Psi}%
_{\sigma }\left( x\right)$ under the transformation Eq.(\ref{TRS}) is
changed to be
\begin{equation}
\hat{\Phi}_{\sigma }\left( x+L\right) =\hat{\Phi}_{\sigma }\left( x\right)
e^{i\sigma _{z}\alpha L},  \label{TBC}
\end{equation}
Correspondingly, the wave function now should satisfy
\begin{equation}
\phi _{\kappa }\left( x_{1},...x_{j}+L...,x_{N}\right) =e^{i\sigma
_{j,z}\alpha L}\phi _{\kappa }\left( x_{1},...x_{j}...,x_{N}\right) .
\label{HIntA}
\end{equation}

The model of two-component bosons described by Eq.(\ref{HamF}) is
analytically solvable by BA under periodic boundary
condition \cite{Y. Q. Li}. The eigenstates can be characterized by the
total spin $S$ of the system which varies from $0$ to $N/2$. The ground
state of interacting two-component bosons corresponds to the ferromagnetic
state with $S=N/2$. Now the problem of solving one-dimensional interacting
two-component bosonic gases with SO coupling is simplified as solving the integrable
model of (\ref{HamF}) under the twisted boundary condition of (\ref{HIntA}),
for which we can still obtain exact solutions by the same method as
originally developed in \cite{B. S. Shastry}. The system is solvable by the
same Bethe-type wavefunction as below
\begin{eqnarray}
\phi _{\kappa }\left( x_{1}...x_{N}\right) &=&\sum_{P,Q}\theta \left(
x_{q_{N}}-x_{q_{N-1}}\right) ...\theta \left( x_{q_{2}}-x_{q_{1}}\right)
\nonumber \\
&&\times A\left( Q,P\right) e^{i\sum_{j}k_{p_{j}}x_{q_{j}}},  \label{WFBA}
\end{eqnarray}%
where $Q=\{q_{1},q_{2},...,q_{N}\}$ and $P=\{p_{1},p_{2},...,p_{N}\}$ denote
permutations of $1,...,N$. $\theta \left( x_{2}-x_{1}\right) $
is the step function and $\{k_{j}\}$ represent quasimomenta with $j=1,...,N$. 
$A\left( Q,P\right)$ are coefficients to be determined, which should fulfill
the following relations
\begin{equation}
A\left( Q;...i,j...\right) =Y_{ji}^{ab}A\left( Q;...j,i...\right) ,
\label{CF}
\end{equation}%
where $Y_{jl}^{ab}=\frac{\left( k_{j}-k_{l}\right) P_{q_{a}q_{b}}-ic}{%
k_{j}-k_{l}+ic}$ with $P_{q_{a} ,q_{b}}$ permutating $q_{a}$, $q_{b}$ in $A\left(
...q_{a},q_{b}...,P\right) $. The scattering matrix remains the same as in
the model with periodic boundary conditions. For eigenstate with total spin $%
S=\frac{1}{2}\left( N-2M\right) $ $(0\leq M\leq N/2)$ under the twisted
boundary condition (\ref{HIntA}), we obtain the following BA equations 
\begin{equation}
e^{i\left( k_{j}-\alpha \right) L}=-\prod_{l=1}^{N}\frac{k_{j}-k_{l}+ic}{%
k_{j}-k_{l}-ic}\prod_{\beta =1}^{M}\frac{k_{j}-\lambda _{\beta }-ic^{\prime }%
}{k_{j}-\lambda _{\beta }+ic^{\prime }},  \label{KEqu}
\end{equation}%
\begin{equation}
\prod_{j=1}^{N}\frac{\lambda _{\zeta }-k_{j}-ic^{\prime }}{\lambda _{\zeta
}-k_{j}+ic^{\prime }}=-e^{i2\alpha L}\prod_{\beta =1}^{M}\frac{\lambda
_{\zeta }-\lambda _{\beta }-ic}{\lambda _{\zeta }-\lambda _{\beta }+ic},
\label{SEqu}
\end{equation}%
where $c^{\prime }=c/2$ and $\{\lambda_{\beta}\}$ denote the spin rapidities.
From the above BA equations Eq.(\ref{KEqu}) and Eq.(\ref{SEqu}), one can observe that
the quasimomenta periodically depend on $\alpha $. So we let $\alpha =\alpha _{0}+%
\frac{2\pi n}{L}$ with $\alpha_{0}\in \left[ 0,\frac{2\pi }{L} \right] $
and $n$ being integer. In fact, we need only consider $\alpha _{0}\in \left[ 0,%
\frac{\pi }{L}\right] $. The quasimomentum for $\alpha _{0}\in \left[ \frac{%
\pi }{L},\frac{2\pi }{L}\right] $ can be deduced from $\alpha _{0}\in \left[
0,\frac{\pi }{L}\right] $ by two steps: first by taking $-\alpha _{0}\in \left[
-\frac{\pi }{L},0\right] $ with solution $-k_{j}$ and $-\lambda _{\zeta }$,
second by shifting $\alpha _{0}$ to $-\alpha _{0}+\frac{2\pi }{L}\in \left[ 
\frac{\pi }{L},\frac{2\pi }{L}\right] $ with the solution unchanging.
\begin{figure}[tbp]
\includegraphics[width=3.5in]{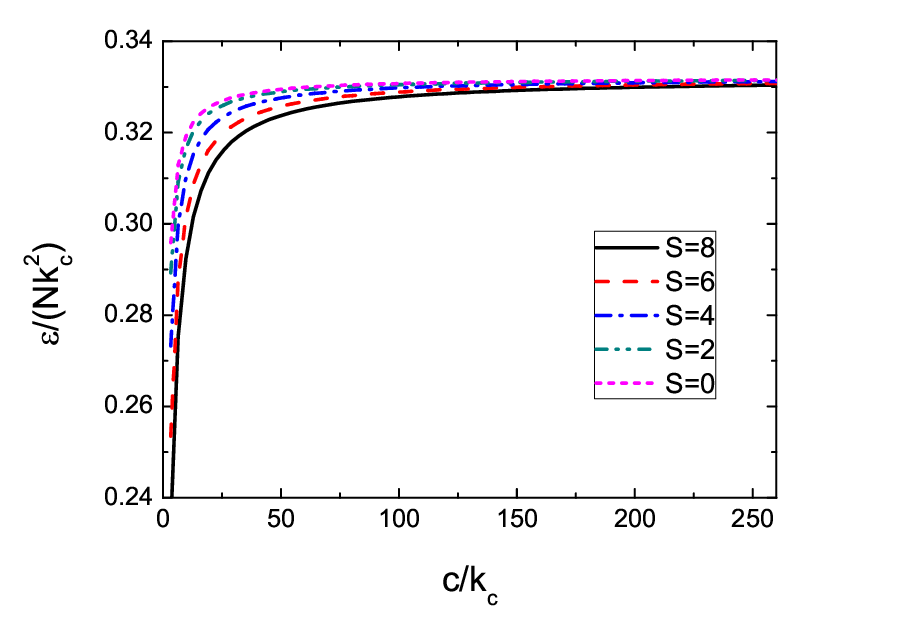}\newline
\caption{(Color online) The lowest energy spectrum as a function of $c$ and the 
total spin $S$ without SOC while $N = 16$, $\alpha =0$. Here $\varepsilon$
is in units of $Nk_{c}^{2}$ and $c$ is in units of $k_{c}$. The ground
state is a ferromagnetic state with $S=8$. These energy levels satisfy $%
E(S)<E(S^{\prime })$ for $S>S^{\prime }$.}
\end{figure}

Taking logarithm of the above equations Eq.(\ref{KEqu}) and Eq.(\ref{SEqu}), we get%
\begin{eqnarray}
\left( k_{j}-\alpha \right) L &=&2\pi I_{j}-\sum_{l=1}^{N}2\tan ^{-1}\frac{%
k_{j}-k_{l}}{c}  \nonumber \\
&&+\sum_{\beta =1}^{M}2\tan ^{-1}\frac{k_{j}-\lambda _{\beta }}{c^{\prime }},
\label{KEQ}
\end{eqnarray}%
\begin{equation}
\sum_{j=1}^{N}2\tan ^{-1}\frac{\lambda _{\zeta }-k_{j}}{c^{\prime }}=2\pi
J_{\zeta }+2\alpha L+\sum_{\beta }2\tan ^{-1}\frac{\lambda _{\zeta }-\lambda
_{\beta }}{c},  \label{SEQ}
\end{equation}%
where $I_{j}$ with $j=1,...,N$ and $J_{\zeta }$ with $\zeta =1,...,M$ denote
the density quantum numbers and the spin quantum numbers, resepectively. Here $I_{j}$ and
$J_{\zeta }$ are integer (half-integer) depending on $N-M$ is odd (even).
Taking the periodic property of $\alpha $ into account and letting $I_{j}=I_{j_{0}}-n$ and $%
J_{\zeta }=J_{\zeta _{0}}-2n$, the above equations could be reduced to
\begin{eqnarray}
\left( k_{j}-\alpha _{0}\right) L &=&2\pi I_{j_{0}}-\sum_{l=1}^{N}2\tan ^{-1}%
\frac{k_{j}-k_{l}}{c}  \nonumber \\
&&+\sum_{\zeta =1}^{M}2\tan ^{-1}\frac{k_{j}-\lambda _{\zeta }}{c^{\prime }},
\label{KEQ0}
\end{eqnarray}%
\begin{eqnarray}
\sum_{j=1}^{N}2\tan ^{-1}\frac{\lambda _{\zeta }-k_{j}}{c^{\prime }} &=&2\pi
J_{\zeta _{0}}+2\alpha _{0}L  \nonumber \\
&&+\sum_{\beta =1}^{M}2\tan ^{-1}\frac{\lambda _{\zeta }-\lambda _{\beta }}{c%
}.  \label{SEQ0}
\end{eqnarray}%
The corresponding eigenenergy is given by
\begin{equation}
E=\varepsilon -N\alpha ^{2}  \label{EE}
\end{equation}%
with $\varepsilon =\sum_{j}k_{j}^{2}$. 
The total momentum is given by
\begin{equation}
K=\tilde{K}-2S\alpha ,  \label{TM}
\end{equation}%
with
\begin{equation}
\tilde{K}=\sum_{j}k_{j}=\frac{2\pi }{L}\left(
\sum_{j=1}^{N}I_{j_{0}}-\sum_{\zeta =1}^{M}J_{\zeta _{0}}\right) +2S\alpha
_{0}.  \label{TMB}
\end{equation}
\begin{figure}[tbp]
\includegraphics[width=3.5in]{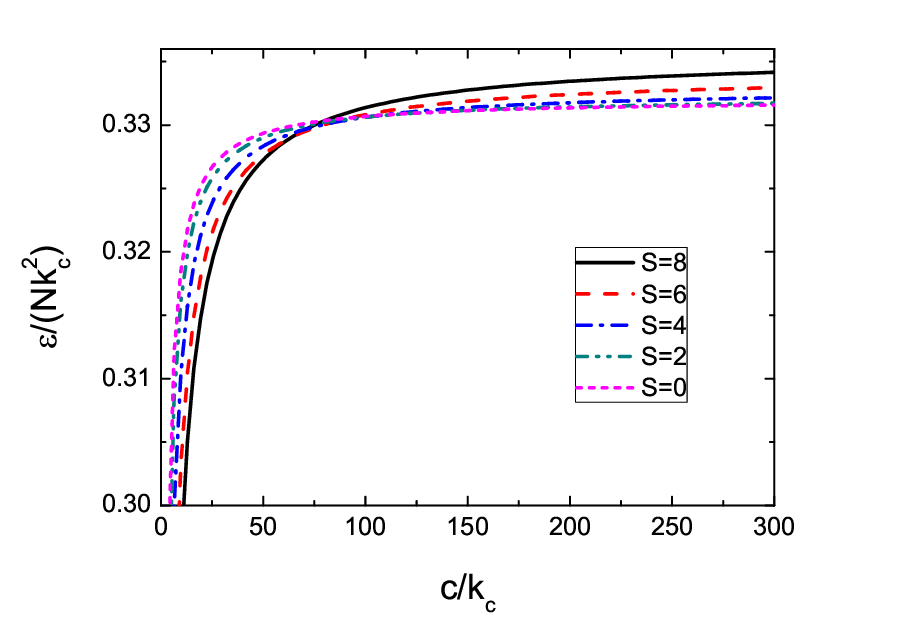}\newline
\caption{(Color online) The lowest energy spectrum versus $c$ for different $S$
with SOC while $N = 16$ and $\alpha =0.95 k_c/N$. The energy
level crossings are shown. In the strong interaction limit, the ground state is transformed to 
the antiferromagnetic state with $S=0$.}
\end{figure}

In the absence of the SO coupling, the model is reduced to the
SU(2) integrable two-component bosonic model \cite{Y. Q. Li,Fuchs,Guan Xiwen}. 
For a given $S$, we can get the eigenenergy by solving BA equations.
The ground state of this system is ferromagnetic state with $S={N}/{2}$ and the 
corresponding ground energy is degenerate for $S_{z}\in \left[ -S,S\right] $. For a system
with $N=16$, we calculate the lowest energies versus the interaction strength $c$
for different total spin $S$, as shown in Fig. 1. Here the parameters $c$ and $E$ are in units
of $k_{c}$ and $Nk_{c}^{2}$ where  $k_{c}$ is defined as $
k_{c}=\pi \rho =\pi {N}/{L}$. Apparently, $E(S)<E(S^{\prime})$ for $
S>S^{\prime }$ in the whole regime of interaction strength and the ground
state is a ferromagnetic state with maximum $S$. This is different from
the case of spin-$1/2$ fermionic gases, where the ground state is antiferromagnetic
state with $S=0$ and $E(S)<E(S^{\prime })$ for $S<S^{\prime
}$ \cite{Lieb-Mattis}. In the strong interaction limit, the ground energy for
different $S$ go to the same value and become degenerate in the
infinite interaction limit \cite{Girardeau07,GuanLM,Deuretzbacher}. 

In the presence of the SO coupling, the energy of ground state 
for different $S$ can be obtained by solving the BA equations. Fig.2 shows 
the lowest energy spectrum as the function of the interaction strength for different §$S$ 
with $\alpha ={0.95k_{c}}/{N}={0.95\pi }/{L}$. For weak interaction,
the energies satisfy $E(S)<E(S^{\prime })$ for $S>S^{\prime}$. When interaction
increases, the level crossing would appear and the ferromagnetic state is no longer
ground state. For the strong enough interaction, the energy levels fulfill the relation of
$E(S)<E(S^{\prime })$ for $S<S^{\prime }$, which is opposite to the law in Fig. 1.
That is to say, the ground state of the two-component bosonic system transfers from the
ferromagnetic state to the antiferromagnetic state.

\section{Strong Coupling Limit}

To futher investigate how the SO coupling affects the ground state energy, we
discuss the strong interaction limit, which permits us to get some
analytical expressions for the energy spectrum. In the strong coupling
limit, $\lambda _{\zeta }$ are proportional to the interaction strength $c$
whereas $k_{j}$ remain finite \cite{GuanLM2}. Applying the Taylor expansion
to Eq.(\ref{KEQ0}) and Eq.(\ref{SEQ0}), the equations of $k_{j}$ and $%
\lambda _{\zeta }$ are simplified as%
\begin{eqnarray}
\left( k_{j}-\frac{\tilde{K}}{N}\right)  &=&\frac{2\pi }{L}I_{j_{0}}-\frac{%
2\pi }{LN}\sum_{j=1}^{N}I_{j_{0}}  \nonumber \\
&&-\frac{2\rho }{c}\left( k_{j}-\frac{\tilde{K}}{N}\right) \left( 1-\frac{v}{%
2N}\right) ,  \label{KSI}
\end{eqnarray}%
\begin{equation}
2N\tan ^{-1}\frac{\lambda _{\zeta }}{c^{\prime }}=2\pi J_{\zeta
_{0}}+2\alpha _{0}L+\sum_{\beta =1}^{M}2\tan ^{-1}\frac{\lambda _{\zeta
}-\lambda _{\beta }}{c}  \label{SSI}
\end{equation}%
with $v=\sum_{\zeta =1}^{M}\frac{1}{1/4+\left( \lambda _{\zeta }/c\right)
^{2}}$. For the state of $M=0$ (or even $M$), $I_{j_{0}}=-\left( N-1\right)
/2...\left( N-1\right) /2$, from Eq.(\ref{KSI}), we can get
\[
\tilde{k}_{j}=\frac{2\pi }{L}I_{j_{0}}-\frac{2\rho }{c}\tilde{k}_{j}\left( 1-%
\frac{v}{2N}\right)
\]%
with $\tilde{k}_{j}=k_{j}-\frac{\tilde{K}}{N}$. For elementary spin
excitations of $M=1$ (odd $M$)\cite{Fuchs}, $I_{j_{0}}^{1}=-N/2,...,N/2-1$,
similarly, $\tilde{k}_{j}$ is given by
\[
\tilde{k}_{j}=\frac{2\pi }{L}\left( I_{j_{0}}^{1}+\frac{1}{2}\right) -\frac{%
2\rho }{c}\tilde{k}_{j}\left( 1-\frac{v}{2N}\right)
\]%
Since $I_{j_{0}}^{1}+\frac{1}{2}=I_{j_{_{0}}}$, in general, the equations
for quasimomenta $k_{j}$ are represented as
\begin{equation}
k_{j}=\frac{2\pi I_{j_{0}}}{L\left( 1+\frac{2\rho }{c}\left( 1-\frac{v}{2N}%
\right) \right) }+\frac{\tilde{K}}{N},  \label{KLI}
\end{equation}%
In the strong coupling limit $c/\rho \rightarrow \infty $, $k_{j}=2\pi
I_{j_{0}}/L+\tilde{K}/N$. From Eq.(\ref{TMB}), it
can be seen that $\tilde{K}$ is influenced by the total spin.
Without SO coupling, the quasimomenta $\left\{k_{j}\right\} $ 
are independent on the total spin. In contrast, with SO coupling, the quasimomenta $\left\{
k_{j}\right\} $ are shifted by the term  $2\alpha _{0}S/N$ for various
total spin $S$. Substituting the above equation into $\varepsilon
=\sum_{j}k_{j}^{2}$, we get
\begin{equation}
\varepsilon =\frac{\pi ^{2}}{3}\frac{N(N^{2}-1)}{L^{2}(1+\frac{2\rho }{c}%
\left( 1-\frac{v}{2N}\right) )^{2}}+\frac{\tilde{K}^{2}}{N}.  \label{ELI}
\end{equation}%
\begin{figure}[tbp]
\includegraphics[width=3.5in]{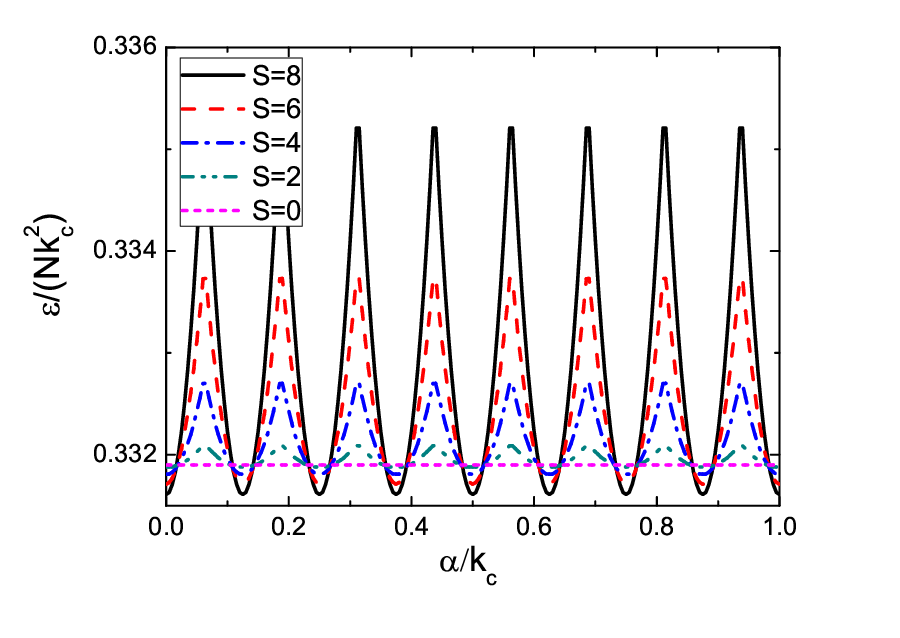}\newline
\caption{(Color online) The lowest energy spectrum as a function of
$\alpha$ for different $S$ with $N=16$, $c/k_{c}=1000$.
The SOC strength $\alpha $ is in units of $k_{c}$
and $\alpha \in (0,k_{c})$. The energy changes cyclically versus $\alpha$.}
\end{figure}
\begin{figure}[tbp]
\includegraphics[width=3.5in]{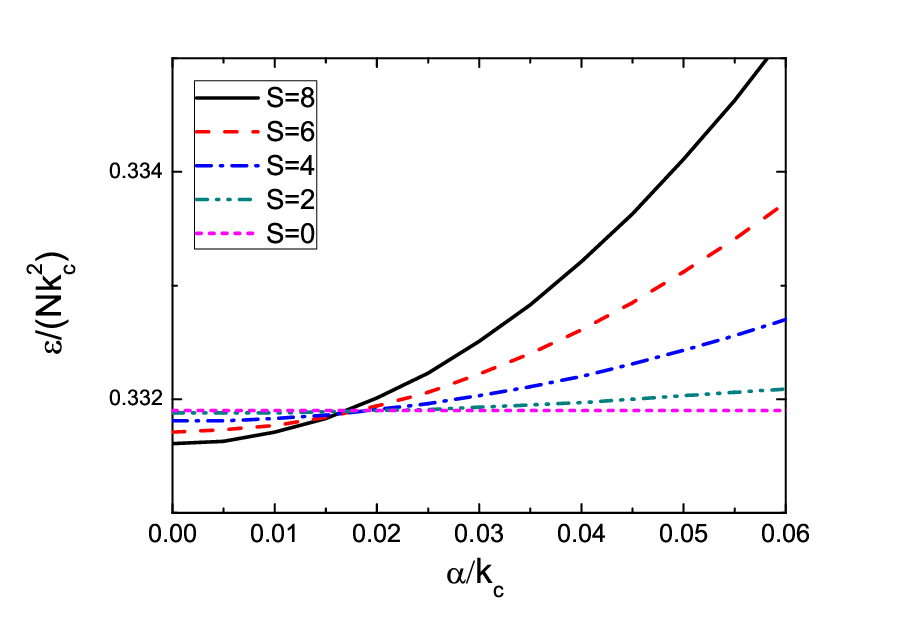}\newline
\caption{(Color online) An enlargement of the energy spectrum shown in Fig.3
for $\alpha \in (0,k_{c}/N)$. The energy crossings are clearly shown. The
ferromagnetic state is the ground state while $\alpha =0$. With the
increase of $\alpha$, the ground state is transformed to the
antiferromagnetic state.}
\end{figure}

From Eq.(\ref{ELI}), it is shown that the ground energy depends on SO
coupling parameter $\alpha $ in two aspects. First, the spin rapidity $%
\lambda _{\zeta }$ is dependent on $\alpha _{0}$ from its self-consistent
Eq.(\ref{SSI}). Second, from Eq.(\ref{TMB}), $\tilde{K}$ is the function of $\alpha _{0}$.
In the strong interaction limit $c/\rho
\rightarrow \infty $, as $\rho /c\rightarrow 0$, the contribution of spin
rapidity can be ignored. Only the term $\tilde{K}^{2}/N$ will affect the
energy for different $S$.

In the case of even $N$ and $M$, the quantum numbers $J_{\zeta
_{0}}=-(M+1)/2-\zeta $ with $\zeta =1...M$. Here from Eq.(\ref{TMB}),
$\tilde{K}=2S\alpha_{0}$ and $\tilde{K}$ is proportional to $S$. From Eq.(\ref{KLI}), while $
c/\rho \rightarrow \infty $, the quasimomenta satisfy $k_{j}=2\pi
I_{j_{0}}/L+2\alpha _{0}S/N$. From Eq.(\ref{ELI}), the energy is given by
\begin{equation}
\frac{\varepsilon }{k_{c}^{2}N}=\frac{N^{2}-1}{3N^{2}}+\left( \frac{\alpha
_{0}}{k_{c}}\right) ^{2}\left( \frac{2S}{N}\right) ^{2}.
\end{equation}%
It is obvious that the energy takes lowest value for $S=0$. 
The SO coupling favors antiferromagnetic state as
the ground state.  
As large but finite interaction, the energy level crossings
occur due to competition between SO coupling and interaction. In Fig. 3, we show the
lowest energy levels versus $\alpha $ in the strongly interacting case with $%
c/k_{c}=1000$, where the SO coupling parameter is chosen in the range of $%
\alpha /k_{c}\in \left( 0,1\right) $. The cyclical change of energy with
the increase of $\alpha $ coincides with our previous discussion. In Fig. 4,
the lowest energy spectrum is plotted versus $\alpha $ with the range
$\alpha /k_{c}\in \left( 0,\pi /L\right)$ which is the half of the period of $\alpha$ in Fig. 3.
When $\alpha =0$, the energy levels fulfill $E(S)<E(S^{\prime })$ for $%
S<S^{\prime }$. With the increase of $\alpha$, energy level crossings appear for different $S$.
When $\alpha $ is large enough, the energy levels satisfy
the opposite law. After the crossing, the energy differences for 
various $S$ become larger with increasing $\alpha$.


Before ending the paper, we would like to give a remark on the interacting
spin-$1/2$ fermionic model with SO coupling, for which the SO coupling
induced level crossing is absent. As for the 1D interacting spin-$1/2$
fermionic gas \cite{CNYang}, the ground state is an antiferromagnetic state (%
$S=0$) in the absence of SO coupling. In the presence of SO coupling, one
can still use the gauge transformation to transform the problem into an
integrable spin-$1/2$ fermionic model with a spin-dependent twisted boundary
condition \cite{Shun Uchino,B. S. Shastry}, and the system is determined by
the following BA equations
\begin{equation}
\left( k_{j}-\alpha \right) L=2\pi I_{j}-\sum_{\zeta =1}^{M}2\tan ^{-1}\frac{%
k_{j}-\lambda _{\zeta }}{c^{\prime }},
\end{equation}%
\begin{equation}
\sum_{j=1}^{N}\tan ^{-1}\frac{\lambda _{\zeta }-k_{j}}{c^{\prime }}=2\pi
J_{\zeta }-2\alpha L+\sum_{\beta =1}^{M}2\tan ^{-1}\frac{\lambda _{\zeta
}-\lambda _{\beta }}{c}.
\end{equation}%
In the strong interaction limit, the energy is
\[
\varepsilon =\frac{\pi ^{2}}{3}\frac{N(N^{2}-1)}{L^{2}\left( 1+\frac{v}{cL}%
\right) ^{2}}+\frac{\tilde{K}^{2}}{N} .
\]%
Similar to the bosonic case, the SO coupling still favors the
antiferromagnetic state as the ground state. The SO coupling does not
lead to level crossing for spin-$1/2$ fermions, i.e., the ground state will
always be an antiferromagnetic state.

\section{Summary}

In summary, we have analytically solved 1D interacting spin-1/2 bosonic
gases with SO coupling. As the effect of SO coupling can be absorbed into
the twisted boundary condition, we get the exact solution to this system by
BA method and find that the corresponding eigenenergies are periodically
dependent on the SO coupling. The interplay between interaction and SO
coupling has revealed the existence of energy level crossing and the ground
state phase transition from the ferromagnetic state to antiferromagnetic
state.

\begin{acknowledgments}
We thank Y. J. Hao for helpful discussions. This work has been supported by
National Program for Basic Research of MOST, NSF of China under Grants
No.11121063 and No.11174360 and 973 grant.
\end{acknowledgments}

\end{document}